# Octave-spanning emission across the visible spectrum from single crystalline 1,3,5,7-tetrakis-(*p*-methoxyphenyl)adamantane

Marius J. Müller,[1] Ferdinand Ziese,[2] Jürgen Belz,[3]
Franziska Hüppe,[3] Saravanan Gowrisankar,[4] Bastian Bernhardt,[4]
Sebastian Schwan,[5] Doreen Mollenhauer,[5] Peter R. Schreiner,[4]
Kerstin Volz,[3] Simone Sanna,[2] and Sangam Chatterjee[1]

[1]*Institute of Experimental Physics I and Center for Materials Research (ZfM/Lama),
Justus Liebig University, Heinrich-Buff-Ring 16, 35392 Giessen, Germany,
sangam.chatterjee@physik.uni-giessen.de*

[2]*Institute of Theoretical Physics and Center for Materials Research (ZfM/Lama),
Justus Liebig University, Heinrich-Buff-Ring 16, 35392 Giessen, Germany, simone.sanna@physik.uni-giessen.de*

[3]*Department of Physics and Materials Science Centre,
Philipps-Universität Marburg, Hans-Meerwein-Str. 6, 35032 Marburg
kerstin.volz@physik.uni-marburg.de*

[4]*Institute of Organic Chemistry, Justus Liebig University, Heinrich-Buff-Ring 17, 35392 Giessen, Germany, prs@uni-giessen.de and Center for Materials Research (ZfM),
Justus Liebig University, Heinrich-Buff-Ring 16, 35392 Giessen, Germany*

[4]*Institute of Physical Chemistry, Justus Liebig University, Heinrich-Buff-Ring 17, 35392 Giessen, Germany, Doreen.Mollenhauer@phys.chemie.uni-giessen.de and Center for Materials Research (ZfM),
Justus Liebig University, Heinrich-Buff-Ring 16, 35392 Giessen, Germany*

**Abstract** Sustainable efficient light emitter based solely on elements-of-hope are needed to replace current compounds based on less-abundant materials. Functionalized diamondoids are a potential solution for this challenge as they offer efficient, octave-spanning emission across the visible spectrum in their single-crystalline form. Its large quantum efficiency increases towards higher-than-ambient temperatures to beyond 7 %. The stability beyond 200 °C renders such functionalized diamondoids as sustainable phosphors for LED applications. Detailed structural and theoretical investigations suggest a crucial role of exciton states accompanied by structural modifications (self-trapped excitons) in the emission process.

## 1. Introduction

Incandescent white-light sources have been vastly replaced by more energy-efficient solutions. Most prominently, such sources are based on light-emitting diodes, and only a few, mostly scientific niche applications, rely on low-etendue, brilliant nonlinear sources. Several concepts for white-light sources are currently pursued. These include mixing red, green, and blue emitters to produce white light. Such an approach is common for organic LEDs, which list low-cost printing capabilities on flexible substrates to their advantages. The vast majority of white-light sources, however, feature LED-driven phosphors. A broad range of compound materials are used to achieve the desired color impressions at high conversion efficiencies. Many of them unfortunately include rare-earth metals or scarcely available elements.

Halide-perovskites are currently considered as efficient emitters, since they show bright white light emission. This is often attributed to the formation of self-trapped excitons. However, these materials also show comparably efficient ion mobilities[1] that render their potential for long-term stable operation not necessarily self-evident.

Other approaches rely on using nanostructures. Their optical properties are commonly extremely size-dependent. One abundant and highly robust class of carbon nanomaterials are diamondoids. They have been considered as promising materials for applications for robust coatings,[2] drugs,[3] molecular electronics,[4] and electron emission[5–7] amongst others.[8,9] Diamondoid nanostructures have also been reported to show spectrally broad photoluminescence, in particular in the UV range.[10] The optical properties of these materials have been studied in great detail.[11–13] The photoluminescence from individual molecules is commonly attributed to the decay from an excited singlet state into the ground state. The observed broad emission has been ascribed to conformational or configurational changes of the individual molecules upon excitation, as low-excitation density and low-temperature matrix spectroscopy show vibrational progressions.[14]

In crystalline materials, however, the relaxation dynamics are more elaborate than for individual molecules. Here, any geometrical changes in the excited state-structures after absorption of photons will reduce the overall symmetry. This can lead to the formation of self-trapped excitons.[15] This mechanism has been proposed earlier for lower diamondoids emitting in the near-UV.[10] The origin of the white-light emission of functionalized diamondoids in the solid state can be clarified by investigation of highly crystalline model structures. To this end, we synthesized and investigated a functionalized diamondoid that yields suitable size single crystal domains for optical spectroscopy. The crystals exhibit bright white-light emission easily observable by the bare eye. The full width at half-maximum spans more than an octave for external quantum efficiencies in the present range for 325 nm excitation. Absorption spectra calculated within density functional theory (DFT) are in excellent agreement with corresponding measurements and explain the origin of the major spectral features.

## 2. Results and Discussion

### 2.1. Optical Properties and Excitation Dynamics

We selected the functionalized diamondoid 1,3,5,7-tetrakis-(*p*-methoxyphenyl)adamantane (**1**) as a model material for a sustainable phosphor, *i.e.*, a light emitter comprising only so-called elements-of-hope rather than, *e.g.*, rare-earth metals. Its mm-length needle-shaped crystals are transparent, the top and bottom surfaces are readily imaged using bright field illumination. The fundamental optical properties are summarized in Fig. 1.

Linear Absorption

The linear absorption derived from diffuse reflectance spectroscopy using the Kubelka-Munk formalism[16–18] is given as a dashed curve in Fig. 1b. It features a prominent steep slope at higher energies. This yields a band energy separation associated with a direct transition around 3.67 eV when applying Tauc's method adapted to crystalline condensed matter systems[18,19] (see Supporting Information Fig. S1); this value is in excellent agreement with the DFT computations discussed below. The electronic structure of **1** is calculated by the density functional model with B3LYP-D3(BJ)/6-31G(d,p). The resulting HOMO-LUMO energy gap of **1** is 5.7 eV. Time-dependent DFT (TDDFT) calculations at B3LYP-D4/cc-pVDZ level of theory (see Supporting Information, chapter 10) similarly yielded an excitation energy of 5.0 eV for the lowest singlet excitation from the HOMO to the LUMO. Computing the point-symmetrical dimer (**1a**) of **1** takes into account as first-order approximation the packing obtained from our experimental crystal structure. The calculated HOMO-LUMO energy gap of dimer **1a** is reduced to 5.4 eV (see Supporting Information Figs. S10 and S11). The lowest singlet excitation of the dimer structure preferred due to energy consideration (**1b**) has been calculated to be 4.7 eV using TDDFT. This already indicates the trend towards a lower HOMO-LUMO energy gap in the crystal computed at 4 eV as discussed below in more detail.

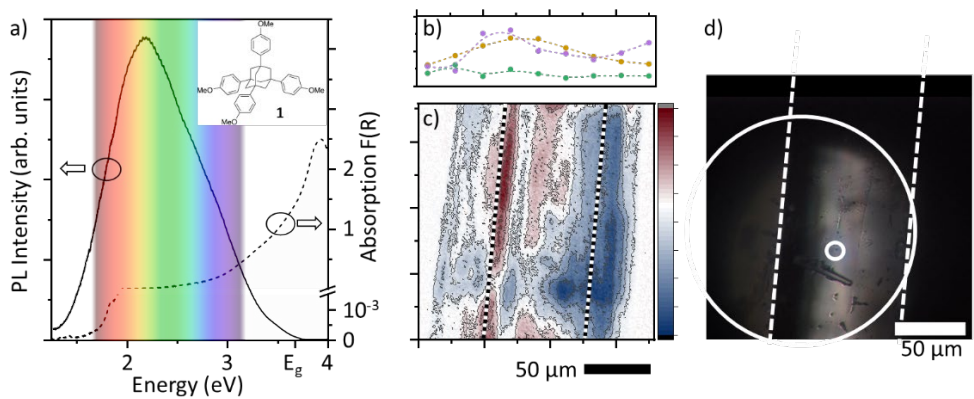

Fig. 1. a) Linear absorption spectrum derived from diffuse reflectance data by the Kubelka-Munk formalism (dashed line) as well as characteristic emission spectra (solid line), b) depth dependence of the three emission peaks, the purple, green, and yellow curves correspond to the areas of the high energy, central, and low-energy fit curves scaled by factors of 100, 4, and 1, respectively. c) Map of the ratio of the surface emission to the bulk emission d) dark-field image corresponding to panel c) the dashed lines in panels c) and d) correspond to each other.

An additional very broadband and weak absorption spans into the near infrared to about 1.25 eV. It is rather featureless across the visible spectral range, *i.e.*, no distinct strongly absorbing resonances are resolved. It only features two changes in curvature which can be associated with indirect transitions around 1.99 and 2.52 eV (Supporting Information Fig. S1). The overall magnitude is estimated to only a few percent from the diffuse reflectance data. This notwithstanding, the samples are very transparent and clear. In the near infrared, two minute peaks at low energies are reproducibly found for several crystals.

Spectrally broad and decaying features below the band gap are often associated with tail states originating from disorder or localized states.[20,21] However, a continuous decaying tail of about 2 eV appears somewhat extraordinary or even peculiar. Alternative explanations are self-trapped excitons, which have been reported to show broad, unstructured emission but can also be observed in the linear response.[22–25] Finally, delocalized excitations in crystals give rise to lower energy absorption features including resonances or continua.[21] These can also be weak, in particular, when associated with dipole-forbidden transitions. Clearly identifying their origin requires in-depth optical and theoretical studies. Fortunately, the high structural quality justifies sophisticated first-principles modeling of the materials properties using periodic boundary conditions in order to support the experimental observations and reveal the consistency of the findings.

## Photoluminescence

The crystals show bright photoluminescence with a blueish-white color impression, which is easily observable by bare eyesight. A characteristic radiometric emission spectrum of the compound for continuous-wave excitation at 3.815 eV (325 nm) under ambient conditions is plotted as a black curve. It covers the complete visible spectrum spanning about 1.5 eV full width at half maximum; it shows a slight asymmetry towards higher energies.

The emission shows several significantly broadened features. These can be well-fitted as sum of three Gaussians and become more evident when studying the temperature dependence of the emission (*cf.* Fig. 2). An exemplary fit is given in Fig. S2. Carefully studying the depth profiles of the emission by scanning the sample vertically through the focus yields clear changes in the spectral line shape. The high-energy emission appears to be more prominent near the sample surface, while the lower-energy features emerge from the bulk material. The areas under the fit curve are plotted versus scan depth in Fig. 1b. The high-energy resonance is scaled by a factor of one hundred. It shows two peaks separated by about 200 – 250 µm depending on the sample thickness. For simplicity, we label the higher-energy emission "surface photoluminescence" as it is most dominant when the laser spot is focused on any of the crystal surfaces. It is observable in virtually all spectra, however, due to the somewhat extended depth-of-focus of the setup. The green curve corresponds to the central emission line and is scaled by a factor of four. It is virtually independent across the depth scan. The low-energy emission (yellow) is the most dominant and, consequently, shown unscaled.

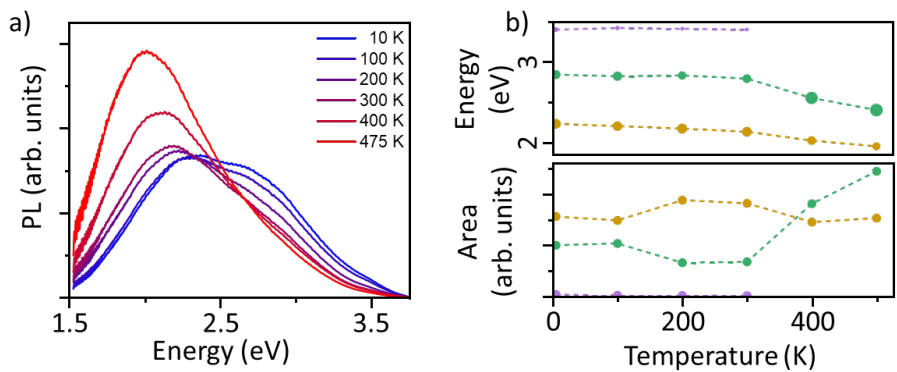

Fig. 2. a) Temperature dependent photoluminescence spectra from 10 K to 475 K for continuous wave excitation at 325 nm b) Temperature dependence of the fit parameters of the deconvolution into three Gaussian contributions. The top panel shows center position energies and the width is indicated as size of the dots. The bottom panel shows the area of the three Gaussians.

The lower-energy emission predominantly emerges from the bulk material and reaches its maximum between the two peaks of the purple line. It is thus termed "bulk emission".

A map of the crystals´ emission reveals the homogeneity of the material and corroborates the depth dependence of the emission. The ratio of the surface emission to the bulk emission is plotted in Fig. 1c in false-colors. Here, the surface emission contribution is strong in regions where the crystal is macroscopically disordered. They correspond well to strong light scattering in the dark-field image given in Fig. 1d.

Investigating the photoluminescence covering the spectral range of the low-energy absorption in more detail aims to clarify the nature of both, the emission and the absorption. Bulk emission spectral spectra for a temperature range spanning from 10 to 475 K and excitation at 3.815 eV (325 nm) are given in Fig. 2a. Intriguingly, the emission increases above room temperature up to lattice temperatures of beyond 200 °C. The area increases by about 40%, boosting external quantum efficiency (see Supporting Information, Fig. S3) from about 5 % at room temperature to above 7 %. This is particularly important for device integration as typical light-emitting diodes or diode laser operate at elevated temperatures. Furthermore, the emission shifts to lower energies with increasing lattice temperatures. This is common for emission from bulk materials as larger lattice constants in crystals commonly reduce the band-gap energies. Hence, the temperature dependence implies an increasing volume of the lattice unit cells with increasing lattice temperatures common to most condensed matter systems.

Above room temperature, the character of the emission changes drastically. Here, the slope of the peak energy shift (*cf.* bottom panel of Fig. 2b) increases at elevated temperatures as does the area of the central peak. The latter even surpasses the contribution of the lowest energy peak. The highest energy component shown in purple virtually vanishes for these temperatures without affecting the goodness of the fit. Notably, the lowest-energy-emission area appears rather independent of the lattice temperature. This hints to a contribution from disorder-related tail states for these photon energies, and, consequently, only a much narrower contribution than the whole tail of 2 eV.

Carrier Temperatures

The excitation density dependence for continuous wave excitation at 2.33 eV (532 nm) is given in Fig. 3a. The exciting laser is eliminated from the spectra using a notch filter (grey-shaded area). This yields both, emission at photon energies above and below the excitation. The spectral shape of normalized emission spectra is very similar on the low-energy side. The high energy side, however, shows significant changes. A detailed analysis of its slope is indicative for the carrier temperature.[20] The ratio of the emission spectra divided by the corresponding absorption yields a measure for the occupation of the density of states accessible by photons. The distribution for quasi-equilibrium and finite temperatures is commonly approximated by a Boltzmann function (see Supporting Information Fig. S4). The derived associated effective carrier temperatures are plotted in Fig. 3b as function of the lattice temperature for various excitation photon fluxes. They are never found to be in equilibrium with the lattice temperatures. This is to be expected as the optical excitation continuously injects carriers with about 0.5 eV of surplus energy assuming a low-energy absorption edge around 1.77 eV (*cf.* Fig. 1a).

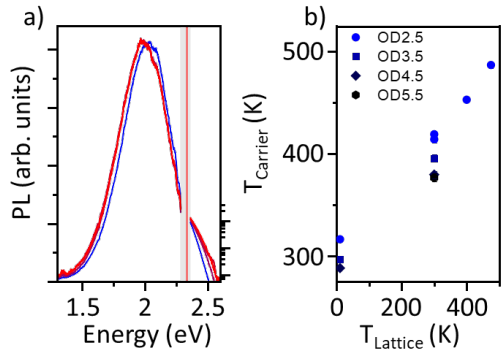

Fig. 3. a) Temperature dependent photoluminescence spectra from 10 K to 475 K for continuous wave excitation at 532 nm b) extracted carrier temperatures by numerically fitting a Boltzmann distribution to the ratio of the high-energy photoluminescence emission shown in the inset of panel a) corrected for the optically accessible density of states inferred from the reflectance data.

The deviation between lattice temperatures and extracted carrier temperatures also increases with decreasing lattice temperatures for one constant excitation photon flux. It varies from 316 to 487 K. Relatively speaking, this means a deviation of 3 160 % for 10 K lattice temperature, which decreases to 103 % for the lattice temperature of 475 K. Such a behavior indicates less efficient cooling of the electronic system for cryogenic lattice temperatures, which is also observed in inorganic semiconductor crystals. This is sometimes attributed to bosonic enhancement, *i.e.*, more efficient phonon emission into populated (phonon) states leading to enhanced cooling rates.[26]

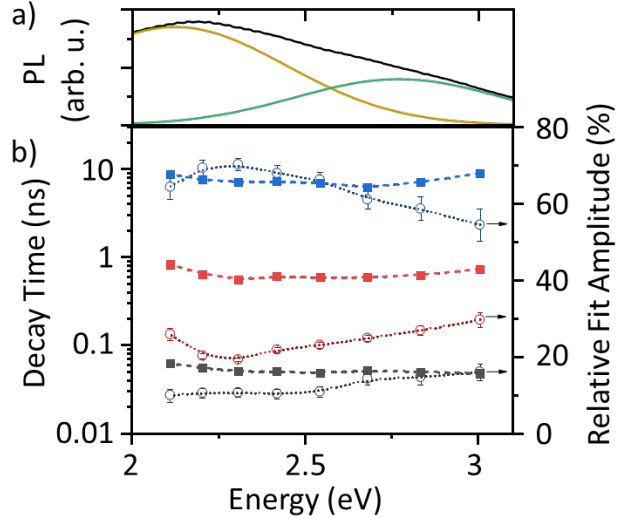

Fig. 4. a) Amplitude fit of the continuous wave PL spectrum in the spectral range where the decay times are measured. b) Lifetimes and relative fit amplitudes of the photoluminescence decay for a 100 fs excitation pulse centered at 380 nm.

Decay Dynamics

The transient photoluminescence can be well reproduced using three decay components. The spectral distribution of the decay times (solid symbols) and the relative weight of the fit amplitudes (open symbols) are summarized in Fig. 4. The photoluminescence spectrum and fit contributions across the experimentally accessible energies is given in Fig. 4a. The fastest decay time component (black) appears virtually constant. Its relative contribution decreases with photon energy across the complete accessible spectral range. Both longer-lived decay times are virtually constant across the energy range covered by the central spectral peak. They then increase for even lower photon energies. The relative weight of the amplitudes changes. The second longest living component increases significantly for the lowest accessible photon energies, virtually eliminating the longest-living component here. The latter becomes less important as the spectral weight of the central peak decreases.

All these observations are consistent with a picture including three groups of excitations. Self-trapped excitons dominate the emission at elevated temperatures, *i.e.*, above room temperature. They are described by the central, extremely broad Gaussian ensemble labelled bulk emission. The high energy contribution is associated with excitations close to the crystal surface, where self-trapping is less favorable as the surface energy tends to inhibit structural changes. The low-energy emission channel is presumably due to tail states, *i.e.*, disorder related emission, as it is less affected by any of the varied parameters.

*2.2. Structural Properties*

Exploring the true microscopic origin of the observations demands establishing a viable concept for the structure-property relationship corroborated by DFT computations. This, in turn, demands understanding the structural properties in great detail, in particular, reconfirming the expected high degree of crystallinity.

The X-ray single crystal structure shows no peculiarities, identifying four molecules per unit cell and the absence of solvent inclusions (Fig. 5a). It is in good agreement to the previously reported structure.[27] Scanning-electron microscopy of the sample's surface shows distinct faceting (*cf.* Fig. 5b). This already hints to the crystalline nature of the sample. The observation of smooth surfaces of several hundreds of square microns reconfirms the feasibility of high-resolution optical micro-spectroscopy on single-crystalline areas.

For transmission electron microscopy (TEM) analysis, two preparation methods were used to obtain a wide range of information: the samples were prepared by focused ion-beam (FIB) milling (Fig. 5b). Comparatively thick (about 400 nm) slices of extremely large lateral dimensions in the range of 20x10 μm$^2$ were cut out of the even larger-scale crystals using a focused ion beam. This preserves the virtually highest possible amount of crystal volume yet inferring inevitable surface damage. Scanning precession electron diffraction provides information on the lateral degree of crystallinity of the specimen. The homogeneous contrast in the virtual dark field image (Fig. 5c) retrieved from one of the diffraction spots in the diffraction pattern shown in the inset confirms the single crystalline nature of the material. Indeed, the sample is single crystal over significantly larger areas as shown by the diffraction patters (Fig. 5d). These were taken from positions all the way across the FIB complete lamella, *i.e.*, around 20 μm.

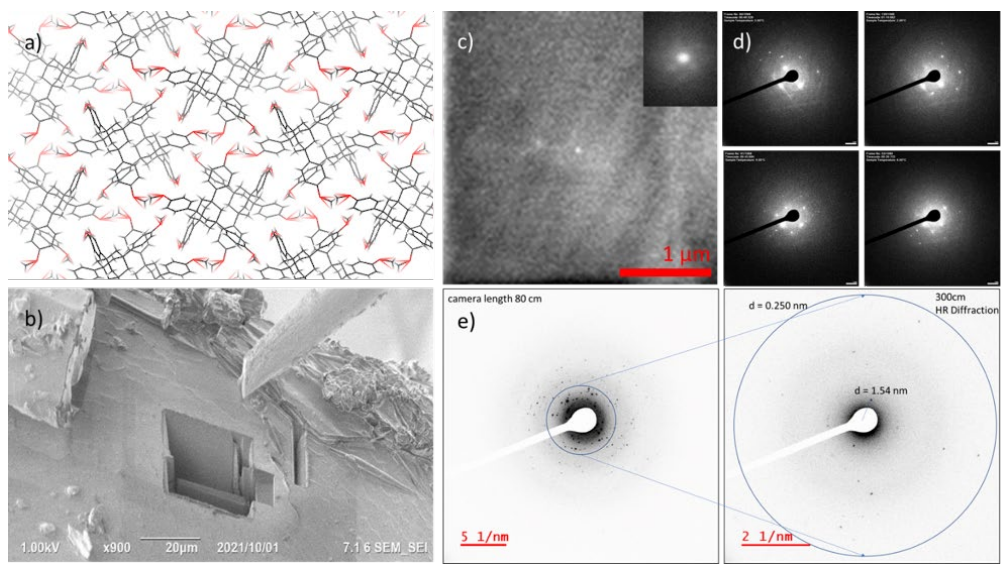

Fig. 5. a) Molecular packing of 1,3,5,7-tetrakis-(*p*-methoxyphenyl)adamantane (1) H atoms omitted for clarity, b) scanning electron microscopic image taken during FIB preparation, c) transmission electron microscopic virtual dark field image retrieved from a scanning precession electron diffraction data set (inset: exemplary diffraction pattern), d) exemplary diffraction patterns taken across the entire lamella (about 20 μm length), e) selected area diffraction patterns taken at different camera lengths.

Material from the same batch was also prepared by ultramicrotomy cutting after embedding it in a resin in order to reconfirm the high quality of the crystals. This technique provides much thinner lamellae without any surface damage. However, the crystals are brittle and easily fracture or even break into small pieces. Two selected area diffraction patters (Fig. 5e), taken at two different camera lengths of the microscopes underline the high degree of crystalline order in the specimen: diffraction spots of lattice planes separated by more than 1.5 nm are found (Fig. 5e). Summarizing, the TEM investigations demonstrate the exceptionally high quality of the synthesized molecular crystals, as well as the large spatial extent of crystalline order.

*2.3 Band Structure*

The high structural quality justifies a description of the investigated compounds using DFT within periodic boundary conditions to describe the experimental observations. In particular, we employ DFT in the independent particle approximation to model the linear optical response of the crystals of **1**. Figure 6 compares the measured (red curve) and computed (black curve) absorption. Both curves show excellent agreement, both in intensity and shape. Interestingly, the computed data are blue-shifted with respect to the measurement, although DFT within (semi)-local xc-potentials is known to underestimate the fundamental electronic gap.[28] This means that quasiparticle effects not included in our models are overcompensated by excitonic effects (electron-hole attraction). This situation is not uncommon in molecular crystals[29,30] and suggests, in turn, large exciton binding energies in crystals of **1**. Including an energy shift of about 400 meV the agreement between measurements and calculations becomes perfect, leading to the superimposition of the pronounced absorption peak labeled (1). Higher-energy resonances, labeled by (2) and (3), are observed in theory, which are experimentally inaccessible due to limitations in the available instrumentation for the medium and vacuum UV.

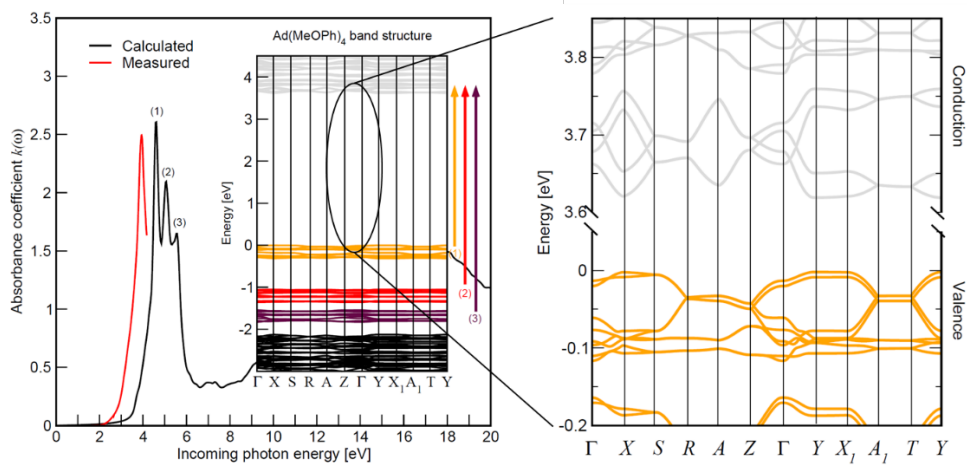

Fig. 6. Absorption spectrum and electronic band structure (inset) calculated for crystalline **1** within PBE in a plane wave basis in the independent-particle approximation. Measured absorption (red curve) for comparison. (1), (2), and (3) label both sub-bands of the valence band as well as electronic transitions from the valence sub-bands to the conduction band. The right-hand panel magnifies the near band-gap region of the band structure.

The electronic band structure presented in Fig. 6 allows to assign the features in the absorption spectrum to electronic transitions from sub-bands in the valence regions to the conduction band. As the sub-bands are separated from electronic gaps, they give rise to close, yet distinct absorption peaks. A drastic drop of the absorbance is predicted for incoming photon energies above 5.5 eV. This seems at first contradicting the calculated band structure, which shows deeper electronic states, cf. the black bands in the inset of Fig. 6. However, the real space representation of the electronic states involved in the transitions explains the absorption spectra.

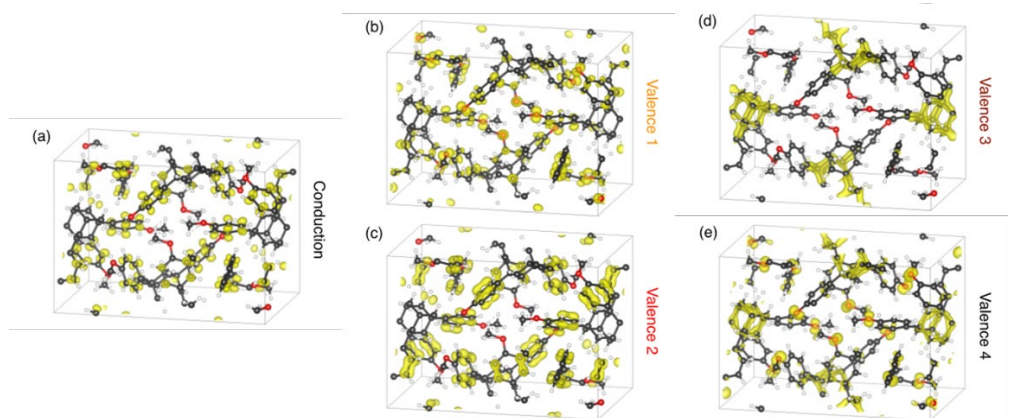

Fig. 7. Absolute squared wavefunctions associated with different electronic states of crystalline **1** computed by PBE in a plane wave basis. (a) Electronic states up to 4 eV above EF, representing the conduction band bottom. (b) to (e): Valence sub-bands represented in orange, red, brown, and black in Fig. 6. An iso-value of 0.02 eV Å$^{-3}$ is shown.

Figure 7 shows the absolute squared wave functions associated with different states. Panel (a) shows the conduction band bottom (up to 4 eV above the Fermi energy). It is clearly localized at the C atoms of the substituent rings, and is reminiscent of the C *p*-orbitals. Panels (b) and (c) show the states associated with the band groups (1) and (2), represented in orange and red in the band structure, respectively. They are also localized at the C atoms of the substituent rings (although (1) shows some contribution from the O atoms) suggesting a high efficiency of the corresponding transitions. Panels (d) and (e) show the states associated with the band groups (3) and the lower lying bands. Both are mostly localized at the diamandoid core. However, while states of the band group (3) have a strong contribution of the C atom of the ligand ring close to the diamandoid core, the energetically deeper states are only localized at the diamandoid cage.

The difference is illustrated in more detail in Fig. 8, which shows the electronic charge in a plane containing two C-C bonds between core and ligands. Thus, transitions from the sub-band (3) to the conduction band are possible, while transitions from the core states to the ligands are hindered from the spatial separation of the involved orbitals. We observe, however, that transitions from (3) to the conduction band have a reduced probability as compared to transitions from (1) and (2) due to the smaller overlap, and therefore feature (3) in the computed spectra is less intense than (1) or (2).

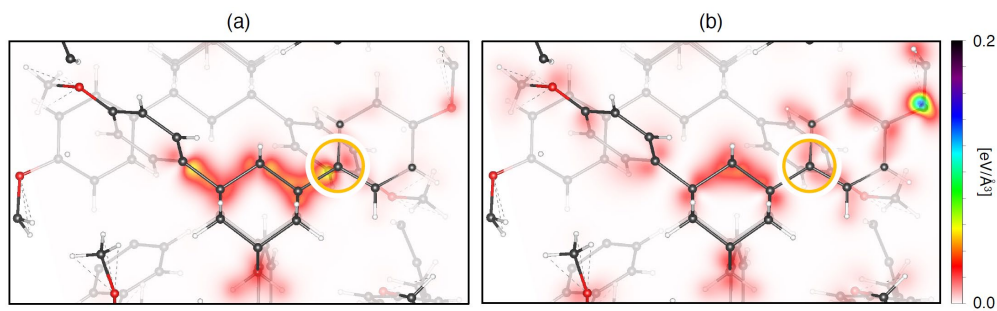

Fig. 8. Electronic charge density associated with the valence sub-bands (3) and (4) calculated for **1** with PBE in a plane wave basis. A plane containing the C-C bond between adamantane core and two p-methoxyphenyl ligands is shown.

On the basis of the DFT results that describe the linear optical response within the single-particle picture, we now consider the implications due to excitonic effects in the measured absorption spectra. The electron-hole interaction is commonly rather strong in molecular crystals, resulting in exciton binding energies of several hundreds of meV. A closer look at the calculated valence and conduction band edges (Fig. 6, right panel) reveals a peculiar feature. Although the electronic states have an overall rather narrow bandwidth, the slope of the band edges is antiphase for the vast majority of the directions in reciprocal space. This results in a multitude of potential ("vertical") optical transitions with similar energies rather than single transition of well-defined energy associated with a high joint density of states and contributions from many regions within the Brillouin zone. Hence, the associated excitons are all embedded in slightly different dielectric environments, depending on their position in real space, which should be accompanied by slightly different degrees of delocalization. This, in turn, would at least contribute to (if not completely explain) the extended low-energy absorption tail as signature of, *e.g.*, dark excitons. The absorption tail may as well contain single-particle contributions, however, excitonic effects are expected to play a crucial role. In particular, if the transition is dipole forbidden, these do not contribute significantly to absorption-like features, unless, selection rules are lifted, for example, by phonon interaction which can lead to charge self-trapping or other symmetry-changing mechanisms.

## 3. Conclusion

In conclusion, we observe an efficient octave-spanning emission across the visible spectrum in a single crystalline functionalized diamondoid. The high structural quality is confirmed by transmission electron microscopy, which resolved lattice plains in the nm-range. Structural changes in the lattice are identified as the origin for broadband spectral emission of the organic crystal. For near-UV excitation the crystals show an intense spectrally broadband emission covering the complete visible spectrum. Its large quantum efficiency of above 5 % even increases towards higher-than-ambient temperatures. The stability beyond 200 °C renders such functionalized diamondoids as sustainable phosphors for LED applications.

## 4. Experimental Section

Synthesis of 1,3,5,7-tetrakis-(p-methoxyphenyl)adamantane (**1**)

A mixture of 1,3,5,7-tetrakis-(*p*-iodophenyl)adamantane (0.8 g) and sodium methoxide (0.6 g) was suspended in a mixture of dry methanol (1 mL) and dry DMF (3 mL). The mixture was stirred and heated at 110 °C for 1 h, after which CuBr (0.61 g) was added in one portion. The reaction was heated and stirred for an additional 6 h and was then allowed to cool to room temperature. The crude product was filtered and extracted with DCM. The solvent was removed by rotary evaporation and the resulting crude solid was dissolved in DCM and subsequently added to methanol to form an off-white solid. Purification via column chromatography (silica gel, ethyl acetate/n-heptane) gave 0.23 g (47%) of **1** as a white solid. **$^{1}$H NMR** (400 MHz, CDCl$_3$) δ = 2.13 (s, 12H), 3.87 (s, 12H), 6.87-6.92 (m, 8H), 7.38-7.41 (m, 8H). **$^{13}$C NMR** (101 MHz, CDCl$_3$) δ = 33.94, 43.00, 50.55, 108.91, 121.29, 137.14, 153.04.

Evaporative crystallization in DCM/n-hexane was utilized to obtain extremely pure crystals of **1** appropriate for single-crystal X-ray diffraction investigation.

Crystal Data and Structure Refinement

Diffraction data for thecrystal structure was collected at 100 K using φ- and ω-scans on a BRUKER D8 Venture system equipped with dual IμS micro-focus sources, a PHOTON100 detector and an OXFORD CRYOSYSTEMS 700 low temperature system. Mo-Kα radiation with a wavelength of 0.71073 Å, Cu-Kα radiation with a wavelength of 1.54178 Å, and a collimating Quazar multilayer mirror were used. Semi-empirical absorption correction from equivalents was applied using SADABS-2016/2[31] and the structures were solved by direct methods using SHELXT2014/5.[32] Refinement was performed against F2 on all data by full-matrix least squares using SHELXL2018/3.[33] All non-hydrogen atoms were refined anisotropically and C–H hydrogen atoms were positioned at geometrically calculated positions and refined using a riding model.

Optical Measurements

Photoluminescence measurements were conducted in a home-built confocal microscope. Resonant excitation experiments were performed using a continuous wave 3.815 eV (325 nm= He-Cd continuous wave laser, sub-bandgap excitation with a single-longitudinal mode, narrow-linewidth continuous wave diode-pumped solid-state laser at 2.33 eV (532 nm). For TRPL measurements, the samples were excited using a pulsed (78 MHz, 100 fs) 380 nm laser. Reflectivity measurements were performed using monochromated light from a Xe-arc lamp and the sample situated under direct illumination in an integrating sphere.

Computational Section

The linear optical response of crystals of **1** were computed with the Vienna ab initio Simulation Package[34,35] utilizing plane-wave DFT. We use an energy cutoff of 450 eV for the expansion of the electronic wave functions in this basis. The electron-ion interaction is efficiently described by projector-augmented-waves (PAW).[36] We employ the popular formulation of the generalized gradient approximation (GGA)[37] proposed by Perdew, Burke and Ernzerhof (PBE). This exchange-correlation (XC) functional has demonstrated numerical accuracy and reliability as well as a high computational efficiency.[30,38] However, as (semi)local XC functionals fail to correctly describe the long-range vdW interactions, we apply the semi-empirical DFT-D3 correction scheme with zero damping,[39] which has been found to be sufficiently accurate in our previous studies.[30,38] A mesh of 4×4×4 k points was employed to sample the Brillouin zone. Starting from the positions as determined by X-ray diffraction (monoclinic supercell of the point group $C_{2h}$), we calculated the imaginary part of the frequency dependent dielectric function by a summation over empty states as proposed by Gajdoš et al..[40] The real part is derived in a second step by means of the Kramers-Kronig transformation. The number of electronic states considered in the computations of the linear optical properties is 2512. Thus, all electronic states within a distance of at least 50 eV from the Fermi energy were included. For the comparison with measured optical spectra we used real and imaginary parts of the dielectric function, $\varepsilon_r(\omega)$ and $\varepsilon_i(\omega)$, respectively. We obtained the absorbance coefficient κ using the approximation

$$\kappa = \sqrt{\sqrt{\varepsilon_r^2 + \varepsilon_i^2} - \varepsilon_r(\omega)}$$

In order to allow for comparison with the experimental data we averaged over the three Cartesian directions to

$$\varepsilon(\omega) = \frac{1}{3} \sum_{i=x,y,z} \varepsilon_{ii}(\omega).$$

**Funding**. Funding through the Deutsche Forschungsgemeinschaft is gratefully appreciated through FOR 2824; S. Chatterjee is supported through the Heisenberg Programme (CH660/08).

**Acknowledgments**. Calculations for this research were conducted on the Lichtenberg high performance computer of the TU Darmstadt and at the Höchstleistungrechenzentrum Stuttgart (HLRS). The authors furthermore acknowledge the computational resources provided by the HPC Core Facility and the HRZ of the Justus Liebig University Giessen.

**Data availability.** Data underlying the results presented in this paper are available in Dataset 1, Ref. [5].

**Supplemental document.** See Supplement 1 for supporting content.